\documentclass[letterpaper,11pt]{article}

\usepackage{amsmath}
\usepackage{amsfonts}
\usepackage{amssymb}
\usepackage{mathrsfs}
\usepackage{multicol}
\usepackage{color}
\usepackage{xcolor}
\usepackage{graphicx}
\usepackage{pstricks}
\usepackage{pst-node}
\usepackage{wrapfig}
\usepackage{enumerate}
\usepackage{caption}
\usepackage[toc,page]{appendix}
\usepackage{cite}
\usepackage{subfigure}
\usepackage{amsfonts,amssymb}
\usepackage{amsthm}
\usepackage{amscd}

\usepackage{hyperref}
\usepackage{url}

\definecolor{gray1}{gray}{0.1}
\definecolor{gray2}{gray}{0.2}
\definecolor{gray3}{gray}{0.3}
\definecolor{gray4}{gray}{0.4}
\definecolor{gray5}{gray}{0.5}
\definecolor{gray6}{gray}{0.6}
\definecolor{gray7}{gray}{0.7}
\definecolor{gray8}{gray}{0.8}
\definecolor{gray9}{gray}{0.9}

\newpsobject{showgrid}{psgrid}{subgriddiv=1,griddots=10,gridlabels=6pt}

\definecolor{dark-green}{rgb}{0,0.7,0}
\definecolor{dark-blue}{rgb}{0,0.2,0.5}
\definecolor{med-blue}{rgb}{0,0.7,1}
\definecolor{mblue}{rgb}{0,0.2,1}
\definecolor{cnc}{rgb}{0.8,0,0}
\definecolor{light-red}{rgb}{1,0.8,0.8}
\definecolor{dark-yelow}{rgb}{1,0.8,0}
\definecolor{light-blue}{rgb}{0.8,0.9,1}
\definecolor{verylight-blue}{rgb}{0.93,0.95,1}
\definecolor{light-yelow}{rgb}{1,0.9,0.8}
\definecolor{grey}{gray}{0.88}

\usepackage{multirow}

\newpsobject{showgrid}{psgrid}{subgriddiv=1,griddots=10,gridlabels=6pt}

\usepackage[normalem]{ulem}
\begin{document}

\thispagestyle{empty}

\setlength{\abovecaptionskip}{10pt}

\begin{center}
{\Large\bfseries\sffamily{Motion of charged particles in an electromagnetic swirling universe: \\
The complete set of solutions}}
\end{center}
\vskip 1cm

\begin{center}
{
\bfseries{\sffamily{Rog\'erio Capobianco$^{\rm 1}$}},
\bfseries{\sffamily{Betti Hartmann$^{\rm 2}$}}, and
\bfseries{\sffamily{Jutta Kunz$^{\rm 3}$}}
}\\
\vskip 0.5cm

{$^{\rm 1}$\normalsize{Instituto de F\'isica de S\~ao Carlos, Universidade de S\~ao Paulo, S\~ao Carlos, S\~ao Paulo 13560-970, Brazil}}\\
\vskip 0.1cm

{$^{\rm 2}$\normalsize{Department of Mathematics, University College London, Gower Street, London, WC1E 6BT, UK }}\\
\vskip 0.1cm
{$^{\rm 3}$\normalsize{Institut f\" ur Physik, Carl-von-Ossietzky Universit\"at Oldenburg, 26111 Oldenburg, Germany}}\\
\end{center}

\vspace{1cm}

\begin{abstract} 
We discuss the motion of electrically and magnetically charged particles in the electromagnetic swirling universe. 
We show that the equations of motion can be decoupled in the Hamilton-Jacobi formalism, revealing the existence of a fourth constant of motion. 
The equations of motion can be analytically integrated.    
The solutions are presented in terms of elementary and elliptic functions. 
In addition, we discuss the possible orbits for both uncharged particles (in which case the motion is geodesic) and charged particles, respectively. 
A typical orbit is bounded in the radial direction and escapes to infinity in the $z-$ direction. 
However, the presence of the electromagnetic fields also leads to the existence of planar orbits. 

\end{abstract}
 
\vspace{1cm}

\centerline{\today}
 
\section{Introduction}

The paradigm of General Relativity is that gravity is described by the space-time curvature. 
In 1968, Ernst developed a new method to solve Einstein's field equation for stationary and axially symmetric space-times \cite{ernst1968new,ernst1968new2}. 
An interesting feature of this new formulation is the simplicity with which the invariance of the field equations can be studied. 
Moreover, it allows to generate new solutions by applying appropriate transformations.
The Harrison transformation allows to embed a seed solution into a magnetized universe.
If, for instance, the seed solution is taken to be the Minkowski space-time, the resulting space-time is the Melvin magnetic universe \cite{ernst1976black}, a static, non-singular, cylindrically symmetric space-time that describes an extended self-gravitating system of an electromagnetic field kept together by its own gravity. 
On the other hand, the Ehlers transformation allows to embed a seed solution into a \textit{swirling universe}, which is a non-singular, stationary space-time that is characterized by its \textit{swirling parameter}. 
This parameter is to be understood as characterizing the background rotation \cite{astorino2022black}. 
Recently, it has been shown that the Harrison and the Ehlers transformations commute, and that interesting new solutions can be generated by the composition of these two transformations \cite{Barrientos:2024pkt,Astorino:2023ifg}. 
For the purpose of this paper, we are interested in a novel solution that is equipped with both the background rotation as well as electromagnetic fields, the so-called \textit{electromagnetic swirling universe} (EMS) \cite{Barrientos:2024pkt}.

A common way to explore space-time's gravitational effects is by studying the dynamics of test particles, which is typically described by a system of coupled second-order ordinary differential equations. 
The space-time symmetries play a fundamental role in this study. 
In particular, there are cases in which the equations of motion can be decoupled and analytically integrated for either geodesic motion or charged particle motion, as was shown for many different space-times, including the Schwarzschild, Reissner-Nordstr\"om and Kerr(-Newman) space-times, respectively \cite{hackmann2013charged,hackmann2010geodesic,Fujita:2009bp,grunau2011geodesics,hagihara1930theory,carter1968global,kerrweierstrass}. 
Furthermore, the geodesic motion in the Melvin space-time can also be analytically integrated \cite{Thorne1965absolute,Melvin:1965zza}, while charged particles have been considered in \cite{Lim:2015oha,Lim:2020fnx}. 
The complete description of the geodesic motion in the swirling universe was also obtained recently \cite{Capobianco:2024swirling}.

In this paper, we study the motion of charged particles in the EMS space-time. 
The general structure of the equations of motion resembles those for geodesic motion in the swirling universe \cite{Capobianco:2024swirling} and accordingly, the equations of motion can be decoupled using the Hamilton-Jacobi formalism, such that a fourth constant of motion can be found. 
The decoupled equations of motion can be analytically integrated using elementary and elliptic functions. 
A typical orbit is bounded in the radial direction and can be planar or escape to infinity in the $z-$direction.

This paper is organized as follows: in Section \ref{EMS space-time}, we introduce the metric tensor and some properties of the space-time. 
In Section \ref{motion in EMS space-time}, we derive the equations of motion. 
The full analytical solutions are presented in Section \ref{section analytical solutions}, where, in addition, we discuss the possible orbits. 
We conclude in Section \ref{conclusions}. 
Throughout this paper, we use the Einstein convention and geometrized units, such that $G = c = 1$, with $G$ Newton's constant and $c$ the speed of light. 
The metric signature used is $(-,+,+,+)$.

%
%
%
%
\section{The electromagnetic swirling universe}
\label{EMS space-time}

The electromagnetic swirling (EMS) space-time is a novel solution found in \cite{Barrientos:2024pkt}. 
The solution is obtained by a composition of Harrison and Ehlers transformations using the Minkowski space-time as seed. 
It describes a rotating background equipped with external electric and magnetic fields. 
The line element describing this space-time reads \cite{Barrientos:2024pkt}
\begin{equation}
\label{Metric EMS}
    {\rm d}s^2 =  \frac{\rho^2}{\Lambda(\rho)} \left( {\rm d}\phi + \omega(z) {\rm d}t \right)^2 + \Lambda(\rho) \left( -{\rm d}t^2 + {\rm d}\rho^2 + {\rm d}z^2  \right), 
\end{equation}
where $\Lambda(\rho) = V(\rho)^2 + j^2 \rho^4$, $V(\rho) =  1 + \vert N \vert ^2 \rho^2 $, $\omega(z) = 4 j z$ and $\vert N\vert^2 =(E^2+B^2)/4$. 
This space-time possesses some interesting discrete symmetries: 
next to invariance under $\{ t,\phi \} \rightarrow \{ -t, -\phi \}$, the metric tensor is also invariant under $\{ t,z \} \rightarrow \{ -t,-z \}$ and $\{ \phi,z\} \rightarrow \{ -\phi,-z \}$. 
Note that the space-time is free of closed timelike curves \cite{Barrientos:2024pkt}.
The gauge field associated to this space-time reads
\begin{equation}
\label{gauge field}
    A = A_\mu {\rm d}x^\mu = \frac{\rho^2}{2\Lambda} \left( 2z \left[ \frac{E V(\rho)^2}{\rho^2} +j \left(2 B V(\rho) - j E \rho^2 \right) \right] {\rm d}t + \left( B V(\rho) - j E \rho^2 \right) {\rm d}\phi \right).
\end{equation}
Note that, the complex-valued metric parameter $2N = E + \imath B$ encodes the information on the electric ($E$), and magnetic ($B$) fields of the background seed solution, the Melvin universe. 
In the following, we will use $\vert N \vert^2 \equiv N^2$ to simplify the notation.

The EMS space-time is of Petrov type D and has no coordinate singularity \cite{Barrientos:2024pkt}. 
In the absence of the swirling parameter, we recover the usual Melvin universe, and for vanishing gauge fields, the space-time corresponds to the swirling universe. 
Therefore, the space-time can be understood as a \textit{Melvin universe} immersed into a swirling universe, or vice-versa. 
The space-time is stationary and possesses an ergoregion for $j\neq 0$ and
\begin{equation}
    \vert 4j \rho z \vert > 1 + 2 N^2 \rho^2 + \left( N^4 + j^2 \right) \rho^4 \ .
\end{equation}

\begin{figure}[p]
    \begin{minipage}{8cm}
        \centering
        \includegraphics[scale=0.45]{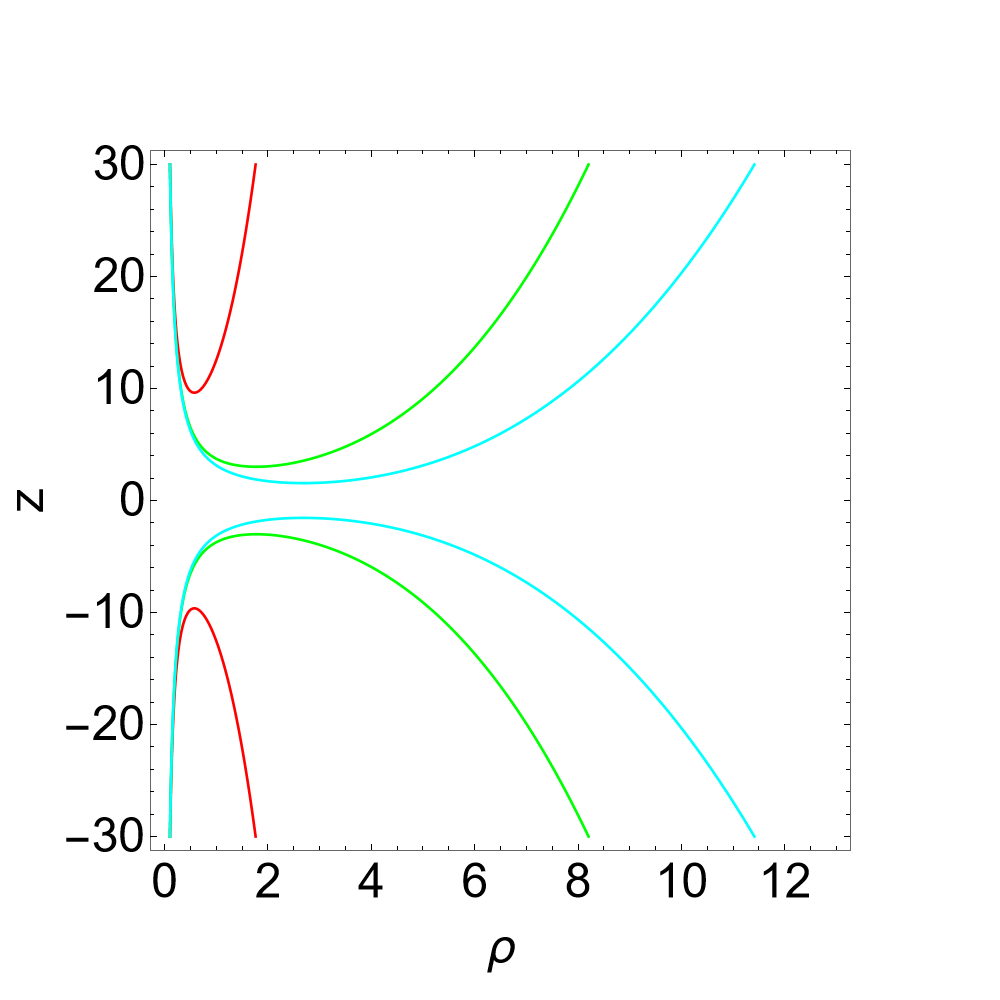}
        \includegraphics[scale=0.4]{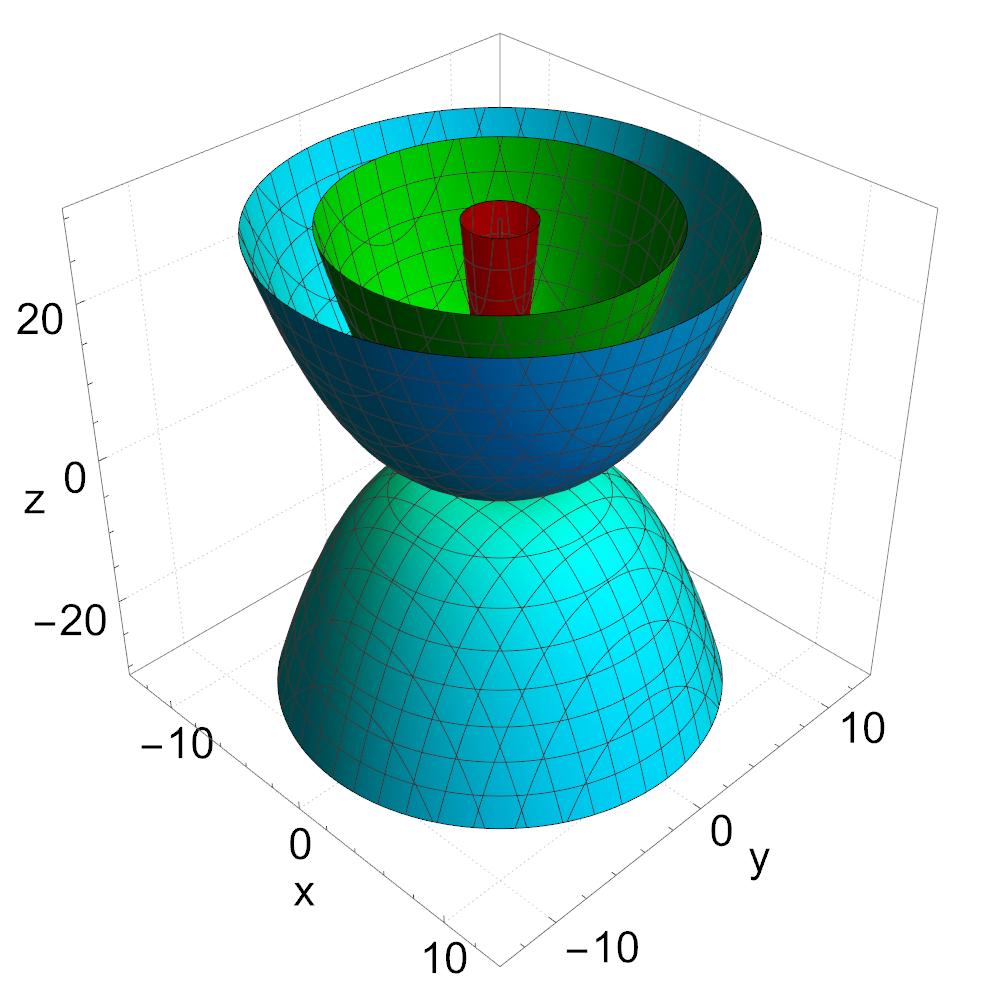}
        \includegraphics[scale=0.8]{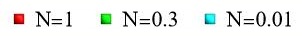}
    \end{minipage}
    \begin{minipage}{8cm}
        \centering
        \includegraphics[scale=0.45]{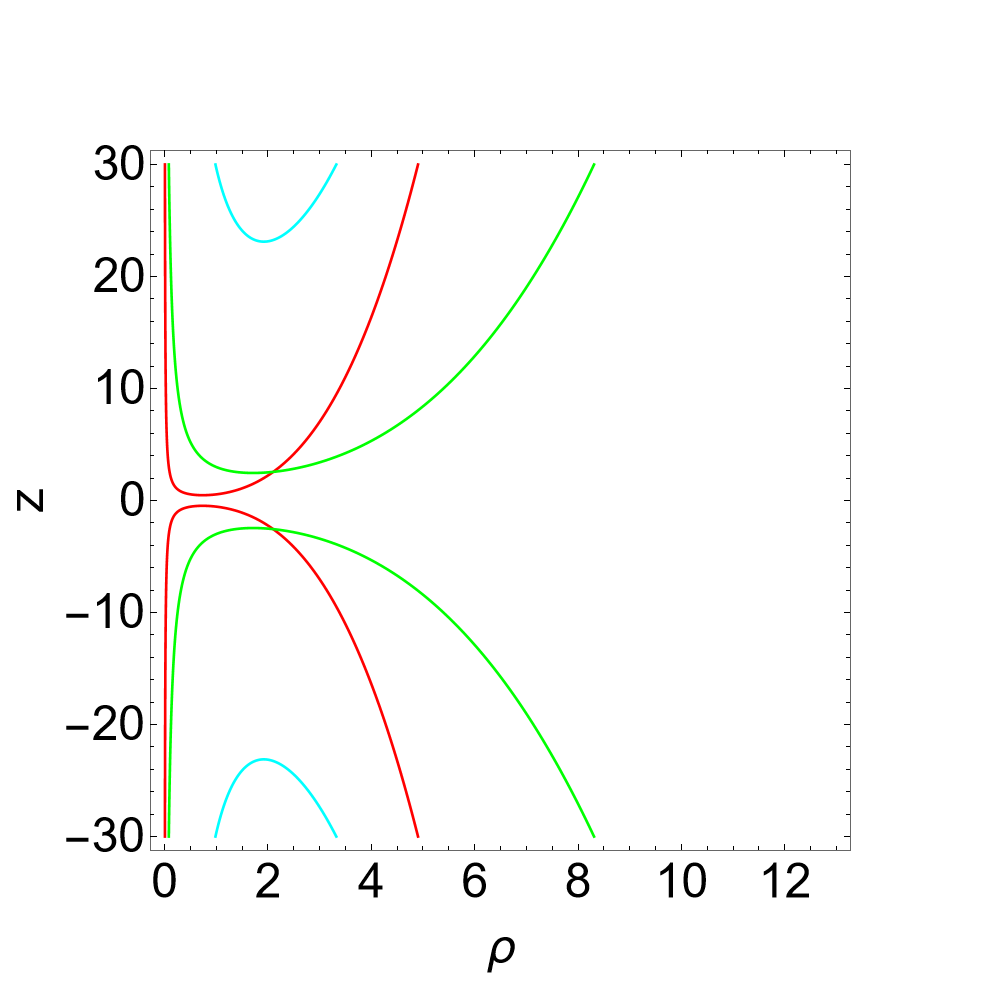}
        \includegraphics[scale=0.4]{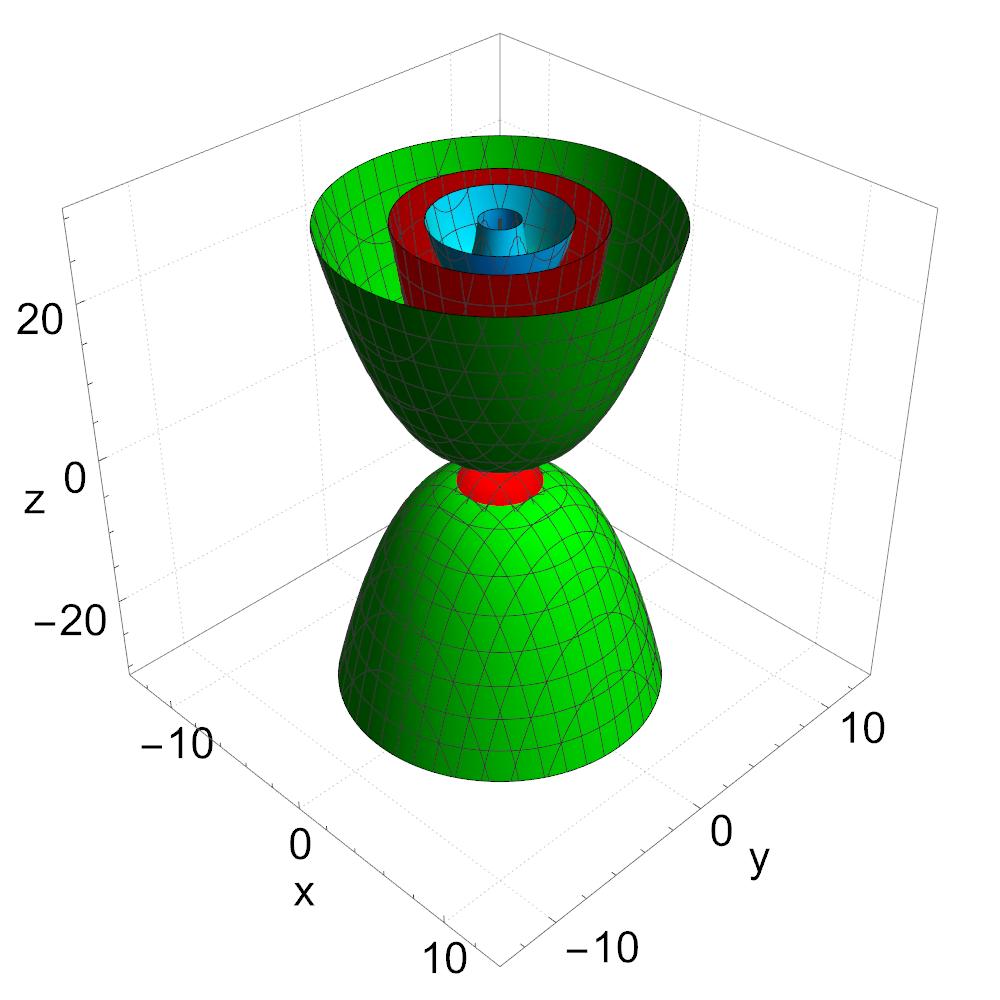}
        \includegraphics[scale=0.8]{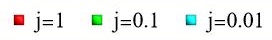}
    \end{minipage}
    \caption{Ergoregions of the electromagnetic swirling universe for different combinations of the metric parameters $N$ and $j$. 
    The plots are shown in the $\rho$-$z$-plane (\textit{upper}), and rotated around the symmetry axes (\textit{lower}), respectively. 
    The \textit{left} figures show the ergoregions for a fixed value of $j=0.08$ and three different values of $N \in \{ 0.01,0.3,1 \}$. 
    The \textit{right} figures show the ergoregions for a fixed value of $N=0.3$ and three different values of $j \in \{ 0.01,0.1,1 \}$.}
    \label{ergoregions_for_EMS}
\end{figure}

The ergoregion resembles, qualitatively, the one of the swirling universe: 
it is composed of two distinct patches, above and below the equatorial plane, extending to infinity. 
The addition of the external electromagnetic field slightly affects the structure as demonstrated in Fig.~(\ref{ergoregions_for_EMS}), where we show the effect by either fixing $j$ and varying $N$ (left) or fixing $N$ and varying $j$ (right). 

%
%
\section{Motion in EMS space-time}
\label{motion in EMS space-time}

In this section, we will describe the motion of test particles in the EMS space-time. For the sake of completeness, we consider a dyonic test particle, i.e., a test particle that is equipped with both electric and magnetic charges. 
Since the space-time described by the metric tensor (\ref{Metric EMS}) and the gauge field (\ref{gauge field}) is stationary and axially symmetric there are two cyclic variables that lead to the existence of two constants of motion,
\begin{eqnarray*}
\label{constant of motion - cyclic variables}
    p_t &=& g_{t t}\dot{t} + g_{t \phi} \dot{\phi} + q A_t -\imath g \Tilde{A}_t := - \mathcal{E}, \\
    p_\phi &=& g_{\phi \phi} \dot{\phi} + g_{\phi t} \dot{t} + q A_\phi -\imath g \Tilde{A}_\phi := L,
\end{eqnarray*}
where $\Tilde{A}_\mu = \imath A_\mu\{(E,B)\rightarrow(-B,E)\}$ and $\mathcal{E}$ is the particle's total energy, related to the time translation invariance, while
$L$ is the angular momentum related to rotational invariance around the $z$-axis.
The dot denotes the derivative with respect to an appropriate affine parameter $\tau$. 
Solving the above system for $\dot{t}$ and $\dot{\phi}$ we find
\begin{eqnarray}
\label{t and phi geodesics}
    \Lambda \dot{t} &=& \mathcal{E} + \left( 4jL + \Delta \right)z,   \\
    \Lambda \dot{\phi} &=& \frac{\Lambda \Bar{L}}{\rho^2} - \omega \left( \mathcal{E} + \left( 4jL + \Delta \right)z \right),
\end{eqnarray}
where we have defined the abbreviation $\Bar{L}=\Bar{L}(\rho)$ as follows
\begin{eqnarray}
\label{redefining angular momentum}
\Bar{L} = \Lambda \left( L - q A_\phi + \imath g \Tilde{A}_\phi \right) = L + \left( 2N^2L + \frac{\Tilde{\Delta}}{2} \right)\rho^2 + \left( (N^4 + j^2)L + \frac{\Tilde{\Delta}N^2 + j\Delta}{2} \right)\rho^4,
\end{eqnarray}
with $\Delta = qE-gB$ and $\Tilde{\Delta}=-qB-gE$. 
Note that, in the absence of electric and magnetic fields, $E=B=0$, one has $\Bar{L}=\Lambda L$, and we recover the motion in the usual swirling universe. 
The electric and magnetic fields appear, in fact, only in $\Delta$ and $\tilde{\Delta}$.
Therefore, these quantities will be used to characterize the motion of charged particles. 
Next to $\mathcal{E}$ and $L$ a third constant of motion is given by the normalization condition
\begin{equation}
\label{constant of motion - normalization}
    g_{\mu\nu} \dot{x}^{\mu} \dot{x}^{\nu} = \chi \ ,
\end{equation}
where $\chi = 0$ for massless particles and $\chi = -1$ for massive particles. 
Note that, using this convention, the velocity field, $\dot{x}^\mu$, and the charges, $q$ and $g$, are given in terms of the particle's rest mass.
A fourth constant of motion can be found by considering the Hamilton-Jacobi equation,
\begin{equation}
\label{Hamilton Jacobi equation}
    2 \frac{\partial S}{\partial \tau} = g^{\mu\nu} \left( \partial_\mu S - q A_\mu +\imath g \Tilde{A}_\mu \right) \left( \partial_\mu S - q A_\nu +\imath g \Tilde{A}_\nu \right),
\end{equation}
where $S$ is the Hamilton principal function. 
If the $\rho$- and $z$-motion can be separated, $S$ may be expressed as follows
\begin{equation}
\label{Hamilton-Jacobi Ansatz}
    S = \frac{1}{2}\chi \tau - \mathcal{E} t + L \phi + S_\rho +S_z.
\end{equation}

In fact, by inserting the above Ansatz into the Hamilton-Jacobi equation (\ref{Hamilton Jacobi equation}) one can verify the separability, which leads to the relations
\begin{eqnarray}
    \rho^2 \left( \partial_\rho S_\rho \right)^2 &=& \left( k + \Lambda \chi \right)\rho^2 - \Bar{L}^2 := R(\rho) \ , \label{Srho equation}
    \\
    \left( \partial_z S_z \right)^2 &=& -k + \left[ \mathcal{E} + \left( 4jL + \Delta \right) z \right]^2 := \Xi(z) \ , \label{Sz equation}
\end{eqnarray}
where $k$ is the separation constant, itself a constant of motion, akin to the Carter constant in the Kerr(-Newman) space-time \cite{carter1968global}.

The equations of motion can then be found by the usual method of setting the derivatives of the function $S$ with respect to the four constants of motion $\mathcal{E}$, $L$, $\chi$, and $k$, to zero.
Instead of using this approach, we consider the relations $\partial_\rho S = p_\rho = \Lambda \dot{\rho}$ and $\partial_z S= p_z =  \Lambda \dot{z} $ and rewrite the four equations of motion for a test particle in the form
\begin{eqnarray}
\label{rho geodesic}    \frac{{\rm d}\rho}{{\rm d}\lambda} &=& \xi_\rho \frac{\sqrt{R(\rho)}}{\rho}, \\
\label{zet geodesic}    \frac{{\rm d} z}{{\rm d}\lambda} &=& \xi_z \sqrt{\Xi(z)}, \\
\label{time geodesic}    \frac{{\rm d} t}{{\rm d}\lambda} &=& \mathcal{E} + \left( 4jL + \Delta \right)z, \\
\label{phi geodesic}    \frac{{\rm d}\phi}{{\rm d}\lambda} &=& \frac{\Lambda \Bar{L}}{\rho^2} - 4jz \left[ \mathcal{E} + \left( 4jL + \Delta \right)z \right],
\end{eqnarray}
where we have introduced the Mino time ${\rm d}\lambda = \Lambda^{-1} {\rm d}\tau$. Here, the quantities $\xi_\rho$ and $\xi_z$ can assume two different values, namely $\pm 1$. Note that these values can be chosen independently for a given orbit, but must then be kept fixed during a calculation.

\section{Full set of solutions}
\label{section analytical solutions}

The equations of motion of a charged particle~(\ref{rho geodesic})-(\ref{phi geodesic}) are structurally similar to the geodesic equations in the swirling universe \cite{Capobianco:2024swirling} and can therefore be solved by a similar procedure. 
The solutions are given in terms of the four constants of motion: 
the particle's energy, $\mathcal{E}$, its angular momentum, $L$, the normalization condition $\chi$, and the Carter constant, $k$. 
Physical orbits are only possible for $R(\rho)\geq 0$ and $\Xi(z)\geq 0$. 
The equations of motion can be analytically integrated using elementary functions as well as the Weierstrass $\wp-$, $\zeta-$, and $\sigma-$ functions. 
Here, the solutions are presented in terms of the Mino time $\lambda$, while the full solution of $\lambda$ in terms of the affine parameter $\tau$ is presented in Appendix \ref{affine}. 
\vspace{0.3cm}

\subsection{\texorpdfstring{$\rho-$}-motion}
\label{section rho motion}

The motion in $\rho-$direction is described by Eq.~(\ref{rho geodesic}) and reads
\begin{equation}
\label{rho motion}
    \rho^2 \left( \frac{{\rm d}\rho}{{\rm d}\lambda} \right)^2 = -\Bar{L}^2 + \left( k + \Lambda \chi \right) \rho^2 := R(\rho) =  \sum_{n=0}^{4} a_n \rho^{2n},
\end{equation}
where the coefficients are given by
\begin{eqnarray}
\label{rho coefficients}
    a_0 &=& -L^2, \qquad a_1 = k + \chi - 4L^2N^2 -L \Tilde{\Delta}, \nonumber \\ 
    a_2 &=& 2N^2\chi -2L^2 (j^2 + 3N^4) -L (3N^2\Tilde{\Delta} + j\Delta) -\frac{\Tilde{\Delta}^2}{4}, \nonumber \\
    a_3 &=& -\frac{1}{2}(4LN^2 + \Tilde{\Delta})\left( 2L (N^4+j^2) + N^2 \Tilde{\Delta} + j \Delta \right) + (j^2 + N^4)\chi, \nonumber \\
    a_4 &=& -\frac{1}{4} \left( 2L (N^4+j^2) + N^2 \Tilde{\Delta} + j \Delta \right)^2. \nonumber
\end{eqnarray}
The above Eq.~(\ref{rho motion}) is bi-quadratic in $\rho$ and can hence be put into the following form
\begin{equation}
\label{rho biquadratic}
    \left( \frac{{\rm d}y}{{\rm d}\lambda} \right)^2 = \sum_{n=0}^{4} \tilde{a}_n y^{n} = P(y),
\end{equation}
where $y=\rho^2$ and $\tilde{a}_n = 4a_n$.

The mathematical structure of Eq.~(\ref{rho biquadratic}) is the same as that of other equations of motion describing radial motion in space-times such as the Reissner-Nordstr\"om space-time, the Kerr(-Newman) space-time 
\cite{grunau2011geodesics, hackmann2013charged,carter1968global,kerrweierstrass,hackmann2010geodesic} and the swirling universe \cite{Capobianco:2024swirling}. 
To solve Eq.~(\ref{rho biquadratic}) consider the transformation $y-y_0 = u^{-1}$, where $y_0$ is a root of $P(y)$. This reduces the order of the polynomial $P$ from four to three and we get
\begin{equation}
    \left( \frac{{\rm d}u}{{\rm d}\lambda} \right) = P_3(u) = \sum_{j=0}^{3} b_j u^j,
\end{equation}
which is cast into Weierstrass form by considering the usual transformation
\begin{equation}
    u = \frac{1}{b_3}\left( 4v - \frac{b_2}{3} \right).
\end{equation}
Hence Eq.~(\ref{rho motion}) reduces to
\begin{equation}
\label{radial geodesics -- weierstrass form}
    \left( \frac{{\rm d}v}{{\rm d}\lambda} \right)^2 = P_W(v) = 4v -g_2 v - g_3,
\end{equation}
where the constants $g_2$ and $g_3$ are
\begin{equation}
\label{Weierstrass invariantes}
        g_2 = -\frac{1}{4} \left( b_1 b_3 - \frac{b_2^2}{3} \right), \qquad g_3 = - \frac{1}{16} \left( b_0 b_3^2 + \frac{2 b_2^3}{27} - \frac{b_1 b_2 b_3}{3} \right).
\end{equation}
Equation (\ref{radial geodesics -- weierstrass form}) allows a solution in terms of the Weierstrass elliptic function $\wp$ in the form
\begin{equation}
    v(\lambda) = \wp(\lambda - \lambda_{in}^{(\rho)}; g_2,g_3),
\end{equation}
where $\lambda_{in}^{(\rho)}$ depends only on the initial conditions, 
\begin{equation}
    \lambda_{in}^{(\rho)} = \lambda_0 + \xi_{\rho} \int_{v_0}^{\infty} \frac{{\rm d} v'}{{\sqrt{P_W(v')}}}, \qquad
    v_0= \frac{1}{4} \left( \frac{b_3}{\rho_0^2 - y_0 } + \frac{b_2}{3} \right), 
\end{equation}
and $\rho_0 = \rho(\lambda_0)$ is the initial value of the radial coordinate. 
Thus, the solution of Eq.~(\ref{rho motion}) is given as follows
\begin{equation}
\label{rho equation: rho of lambda}
        \rho(\lambda) = \sqrt{\frac{b_3}{4 \wp \left( \lambda - \lambda_{in}^{(\rho)}; g_2, g_3 \right) - \frac{b_2}{3}} + y_0} \ .
\end{equation}

\subsection{\texorpdfstring{$z-$}-motion}

The motion in the $z$-direction is governed by
\begin{equation}
\label{zet motion}
    \left( \frac{{\rm d}z}{{\rm d}\lambda} \right)^2 = -k + \left( \mathcal{E} + (4jL + \Delta) z \right)^2.
\end{equation}
Here we distinguish two cases.
For $4jL+\Delta = 0$, the above equation can be trivially integrated, thus
\begin{equation}
\label{solution_z_of_lambda_first_case}
    z(\lambda) = z_0 + \sqrt{\mathcal{E}^2 - k} \left( \lambda - \lambda_0 \right),
\end{equation}
where $z(\lambda_0)=z_0$ is the initial value of the $z-$coordinate. 
Since $z$ must be real, physical orbits also require $\mathcal{E}^2 \geq k$. 

For $4jL+\Delta \neq 0$ and $k>0$ the motion is restricted to
\begin{equation}
    z < z_- := -\frac{\mathcal{E}(4jL+\Delta) + \sqrt{k} \vert 4jL+\Delta \vert}{(4jL+\Delta)^2},
    \ \qquad \ 
    z > z_+ := \frac{-\mathcal{E}(4jL+\Delta) + \sqrt{k} \vert 4jL+\Delta \vert}{(4jL+\Delta)^2}. 
\end{equation}
In that case Eq.~(\ref{zet motion}) can also be directly integrated, 
%
\begin{equation}
    \int_{\lambda_0}^{\lambda} {\rm d}\lambda' = \int_{\tilde{z}_0}^{\tilde{z}} \frac{{\rm d}z'}{\sqrt{1-z'}\sqrt{1+z'}} = \frac{1}{4jL+\Delta}\cosh^{-1} (z')|_{\tilde{z}_0}^{\Tilde{z}}
\end{equation}
where $\tilde{z} = \frac{\mathcal{E}+(4jL+\Delta)z}{\sqrt{k}}$, and hence
\begin{equation}
\label{z of lambda}
    z(\lambda) = \frac{1}{4jL + \Delta} \left(-\mathcal{E} + \sqrt{k}\cosh{\left[(4jL + \Delta)(\lambda - \lambda_{in}^{(z)})\right]} \right),
\end{equation}
where $\lambda_{in}^{(z)}$ depends only on the initial conditions
\begin{equation}
    \lambda_{in}^{(z)} = \lambda_0 - \frac{\xi_z}{4jL + \Delta}\cosh^{-1}\left( \frac{\mathcal{E}+(4jL+\Delta)z_0}{\sqrt{k}} \right).
\end{equation}
\subsection{\texorpdfstring{$t-$}-motion}

The motion in the $t-$direction is described by Eq.~(\ref{time geodesic})
\begin{equation}
    \frac{{\rm d}t}{{\rm d}\lambda}  = \mathcal{E} + \left( 4jL + \Delta \right)z(\lambda),
\end{equation}
which can be integrated by distinguishing again  the cases discussed in the above subsection, 
\begin{eqnarray}
    t(\lambda) = 
    \begin{cases}
        t_0 + \mathcal{E} \left( \lambda - \lambda_0 \right), \quad &\text{if} \quad 4jL+\Delta = 0, \\
        \\
        \frac{\sqrt{k}}{4jL + \Delta}\sinh{\left[(4jL + \Delta)~\left(\lambda-\lambda_{in}^{(z)}\right)\right]} + t_{in} \ , \quad &\text{if} \quad 4jL+\Delta \neq 0,
    \end{cases}
\end{eqnarray}
with $t_{in} = t_0 - \frac{\sqrt{k}}{4jL + \Delta}\sinh{\left[(4jL + \Delta)\left( \lambda_0 - \lambda_{in}^{(z)} \right)\right]} $, and $t_0 = t(\lambda_0)$ is the initial value of the time coordinate. 
\subsubsection{\texorpdfstring{$\phi-$}-motion}

The $\phi$-motion is described by the equation
\begin{equation}
    \frac{{\rm d}\phi}{{\rm d}\lambda}  = \frac{\Lambda \Bar{L}}{\rho^2} - 4jz \left( \mathcal{E} + (4jL+\Delta)z \right).
\end{equation}
This equation can be integrated using the solutions for $\rho(\lambda)$ and $z(\lambda)$ obtained above.
Considering the respective parts separately,
\begin{equation}
\label{phi of lambda}
    \int_{\phi_0}^{\phi} {\rm d}\phi' = \phi(\lambda) - \phi_{0} = \mathcal{I}_\rho - \mathcal{I}_z,
\end{equation}
where $\phi(\lambda_0)=\phi_{0}$ is the initial condition of $\phi$, and the two contributions are
\begin{equation}
    \mathcal{I}_\rho = \int_{\lambda_0}^{\lambda} \frac{\Lambda \Bar{L}}{\rho^2}{\rm d}\lambda, \ \ \ \qquad \ \ \ \mathcal{I}_z = 4j \int_{\lambda_0}^{\lambda} z \left( \mathcal{E} + (4jL+\Delta)z \right) {\rm d}\lambda.
\end{equation}

In the following, these two parts will be considered separately.
First, by inserting the same set of transformations discussed in Sec.~\ref{section rho motion}, and performing a partial fraction decomposition, the integral $\mathcal{I}_\rho$ can be rewritten in the form
\begin{equation}
    \mathcal{I}_\rho = \int_{v_0}^{v} f(v') \frac{{\rm d}v}{\sqrt{P_W(v)}},
\end{equation}
with 
\begin{eqnarray}
\label{elliptic third kind - v form}
    f(v) &=& K_0  +\frac{K_1}{v-\alpha} + \frac{K_2}{(v-\alpha)^2} + \frac{K_3}{(v-\alpha)^3} + \frac{C_1}{v-\beta},
\end{eqnarray}
where $\alpha = \frac{b_2}{12}$ and $\beta = \frac{b_2}{12} - \frac{b_3}{4 y_0}$ are the roots of $D(v) =(b_2-1 2v)^3 (-3b_3 + (b_2-12v)y_0) $, and the coefficients are given by
\begin{eqnarray}
    K_0 &=& \frac{\Lambda(\sqrt{y_0})\left( 2\Lambda(\sqrt{y_0}) L + y_0 \left( \tilde{\Delta}V(\sqrt{y_0})-jy_0 \Delta \right)  \right)}{2y_0}, \nonumber \\
    K_1 &=& -\frac{b_3}{8} \left[ 2L \left( j^4 y_0^2 +N^4 \left(6+ N^2 y_0 (8 + 3N^2 y_0) \right) \right) + 3\tilde{\Delta}N^2(1+N^2y_0) +j\Delta (1+N^2 y_0)(1+3N^2y_0) + \right. \nonumber \\
    && \left. j^2 \left( 4L(1+N^2 y_0)(1+3N^2y_0) + \tilde{\Delta}y_0 (2+3N^2y_0) + 3j y_0^2 \Delta  \right) 
    \right], \nonumber \\
    K_2 &=& \frac{b_3^2}{32}\left[ 2LN^6 (4 + 3N^2y_0) + 3N^4 \tilde{\Delta}V(\sqrt{y_0}) + jN^2 \Delta (2+3N^2y_0) + \right. \nonumber \\
    && \left. j^2 \left( \tilde{\Delta} + N^2 (8L + 3y_0 (4LN^2 +\tilde{\Delta})) \right) + 3j^3 y_0 \Delta +6j^4 L y_0 \right], \nonumber \\
    K_3 &=& \frac{b_3^3\left( N^4 +j^2 \right)}{128}\left( 2L \left( N^4 +j^2 \right) + N^2 \Tilde{\Delta} + j\Delta \right), \ \ \ \qquad \ \ \ C_1= - \frac{b_3L}{4y_0^2}.
\end{eqnarray}

Now, inserting the solution found in Sec.~\ref{section rho motion}, $v\left( \lambda \right) = \wp\left( \lambda-\lambda_{in}^{(\rho)} \right)$, into Eq.~(\ref{elliptic third kind - v form}), we conclude that $\mathcal{I}_\rho$ is an elliptic integral of the third kind and hence can be integrated directly (see Appendix \ref{integration of elliptic integrals of the third kind} for details) to give
\begin{eqnarray}
\label{phi rho component}
    \mathcal{I}_\rho &=& \gamma_0 \left( \lambda - \lambda_0 \right) + 
                \gamma_1 \left( \wp(\lambda - \lambda_{in}^{(\rho)} + y_\alpha) - \wp(\lambda - \lambda_{in}^{(\rho)} - y_\alpha) - \wp(\lambda_0 - \lambda_{in}^{(\rho)} + y_\alpha) + \wp(\lambda_0 - \lambda_{in}^{(\rho)} - y_\alpha) \right)  
                \nonumber \\
                & & + \gamma_2 \left[ \zeta\left( \lambda - \lambda_{in}^{(\rho)} - y_\alpha\right) + \zeta \left( \lambda - \lambda_{in}^{(\rho)} + y_\alpha \right) -\zeta\left( \lambda_0 - \lambda_{in}^{(\rho)} - y_\alpha\right) - \zeta \left( \lambda_0 - \lambda_{in}^{(\rho)} + y_\alpha \right) \right] 
                \nonumber \\
              & &  +  \gamma_3 \left[  \ln \left( \frac{\sigma(\lambda - \lambda_{in}^{(\rho)}- y_\beta)}{\sigma(\lambda - \lambda_{in}^{(\rho)}+ y_\beta)}\right) - 
                \ln \left( \frac{\sigma(\lambda_0 - \lambda_{in}^{(\rho)}- y_\beta)}{\sigma(\lambda_0 - \lambda_{in}^{(\rho)}+ y_\beta)} \right)\right] \nonumber \\
              & & +  \gamma_4  
                    \left[  \ln \left( \frac{\sigma(\lambda - \lambda_{in}^{(\rho)}- y_\alpha)}{\sigma(\lambda - \lambda_{in}^{(\rho)}+ y_\alpha)}\right) - 
                    \ln \left( \frac{\sigma(\lambda_0 - \lambda_{in}^{(\rho)}- y_\alpha)}{\sigma(\lambda_0 - \lambda_{in}^{(\rho)}+ y_\alpha)} \right)\right] \ ,
\end{eqnarray}
where the values of the constants are given by
\begin{eqnarray*}
    \gamma_0 &=& \left[  K_0 + \frac{2 \zeta(y_\alpha) K_1}{\wp'(y_\alpha)} + \frac{2 \zeta(y_\beta) C_1}{\wp'(y_\beta)} - \frac{K_3}{\wp'(y_\alpha)^2} \left( 1 + \frac{12 \wp(y_\alpha) \zeta(y_\alpha)}{\wp'(y_\alpha)} \right) \right.    
    \nonumber \\
    & & \left. + \frac{1}{\wp'(y_\alpha)^2}\left(\wp(y_\alpha) + \frac{\wp''(y_\alpha)\zeta(y_\alpha)}{\wp'(y_\alpha)} \right) \left( \frac{3K_3 \wp''(y_\alpha)}{\wp'(y_\alpha)^2} - 2K_2 \right)
    \right] \ ,
    \\
    \gamma_1 &=& \frac{K_3}{2 \wp'(y_\alpha)^3} \ , \\
    \gamma_2 &=& \frac{1}{\wp'(y_\alpha)^2} \left(-K_2 + \frac{3 K_3 \wp''(y_\alpha)}{2\wp'(y_\alpha)} \right) \ , \\
    \gamma_3 &=& \frac{C_1}{\wp'(y_\beta)} \ , \\
    \gamma_4 &=& \frac{K_1}{\wp'(y_\alpha)} + 
            \frac{K_2 \wp''(y_\alpha)}{\wp'(y_\alpha)^3} - 
            \frac{3 K_3}{ \wp'(y_\alpha)^3} \left( 2 \wp(y_\alpha) - \frac{\wp''(y_\alpha)^2}{2 \wp'(y_\alpha)^2} \right) \ , 
\end{eqnarray*}
and $y_\alpha$ and $y_\beta$ are values of the inverse Weierstrass $\wp$-function, i.e.  $\wp(y_\alpha) = \alpha$ and $\wp(y_\beta) = \beta$. $\zeta(y)$ and $\sigma(y)$ are, respectively, the Weierstrass $\zeta-$ and $\sigma-$function. 
We point out, however, that special attention is required when evaluating the logarithm in Eq.~(\ref{phi rho component}).
In this paper, we have implemented the solutions using \textit{Wolfram MATHEMATICA}.
In order to guarantee a proper choice of the complex branch, and therefore a continuous evaluation of $\phi(\lambda)$, we have used the same strategy as introduced in \cite{kerrweierstrass}.

When considering the second integral $\mathcal{I}_z$, we again need to distinguish two cases.
For $4jL+\Delta = 0$, we see that the integration can be trivially done by inserting the solution Eq.~(\ref{solution_z_of_lambda_first_case}).
We then find
\begin{equation}
\label{phi_z_component_first_case}
    \mathcal{I}_z = 4j\mathcal{E}\left[ z_0 \left( \lambda - \lambda_0 \right) + \xi_z \sqrt{\mathcal{E}^2 - k} \left( \lambda - \lambda_0 \right)^2 \right].
\end{equation}
For $4jL + \Delta \neq 0$ we follow the same procedure, but consider the form Eq.~(\ref{z of lambda}), thus
\begin{eqnarray}
\label{phi_z component}
    \mathcal{I}_z &=& \frac{j\sqrt{k}}{(4jL + \Delta)^2}  \left[
    2\sqrt{k} \left( 4jL + \Delta \right) \left( \lambda - \lambda_0 \right) 
    \right. \nonumber \\ && \left. \hspace*{2.2cm}
    - 4\mathcal{E} \left( \sinh{\left[(4jL + \Delta)(\lambda - \lambda_{in}^{(z)}) \right]} - \sinh{\left[(4jL + \Delta)(\lambda_0 - \lambda_{in}^{(z)}) \right]}  \right) \right. \nonumber \\ 
    && \left. \hspace*{2.2cm}
    + \sqrt{k} \left( \sinh{\left[ 2 \left( 4jL + \Delta \right)\left( \lambda - \lambda_{in}^{(z)} \right) \right]} - \sinh{\left[ 2 \left( 4jL + \Delta \right)\left( \lambda_0 - \lambda_{in}^{(z)} \right) \right]}  \right) \right] \ .
\end{eqnarray}
Finally, Eq.~(\ref{phi of lambda}), together with Eq.~(\ref{phi rho component}) and either Eq.~(\ref{phi_z_component_first_case}) or Eq.~(\ref{phi_z component}), depending on the case of interest, concludes the integration of the $\phi-$equation.  
\subsection{Classification of orbits}

Based on the solutions of the equations of motion presented above we now study the resulting orbits in the EMS space-time. 
The characterization of the orbits can be done in terms of the four constants of motion, namely ${E}$, $L$, $\chi$, and $k$. 
In the following, we will consider both, the charged particle motion and the geodesic motion.

\begin{figure}[p]
    \centering
        \subfigure[ $\chi=0$, $\Delta = 0$, $\tilde{\Delta}=0$.]{\includegraphics[scale=.345]{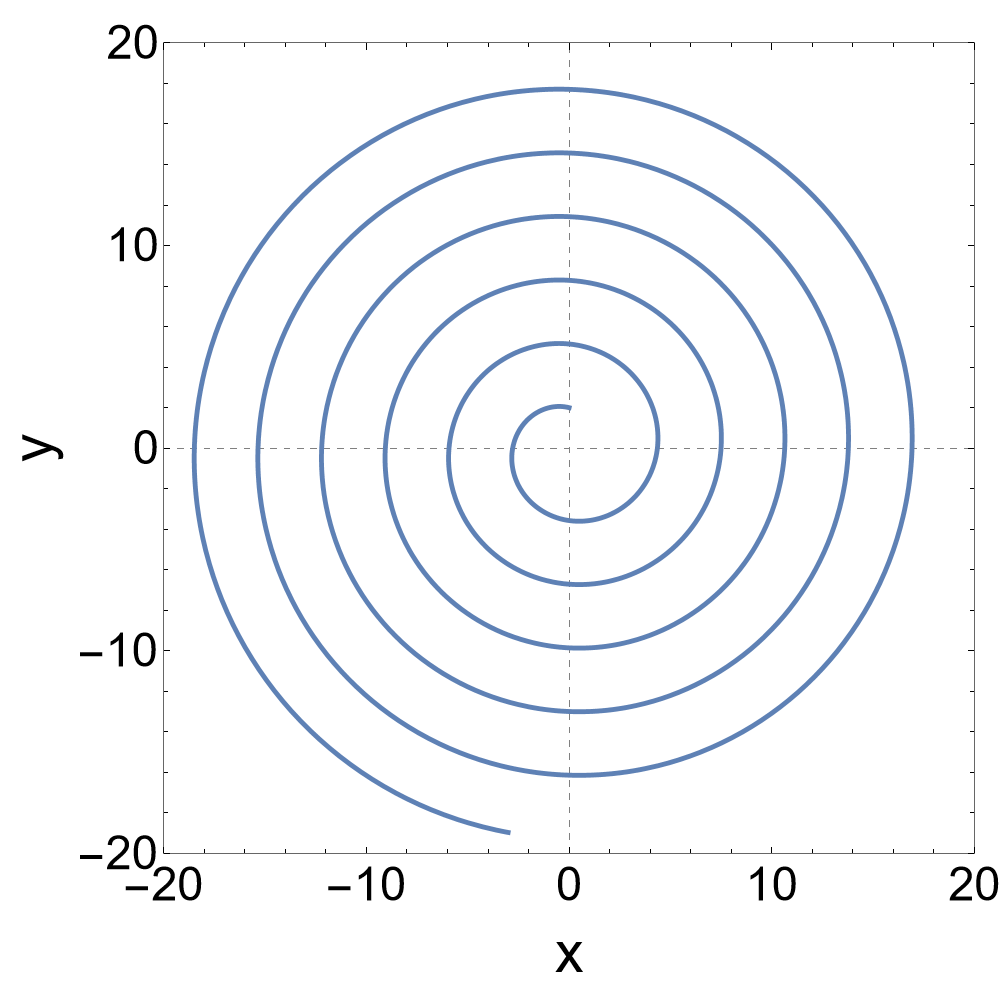}}
        \subfigure[ $\chi=-1$, $\Delta = 0$, $\tilde{\Delta}=0$.]{\includegraphics[scale=.34]{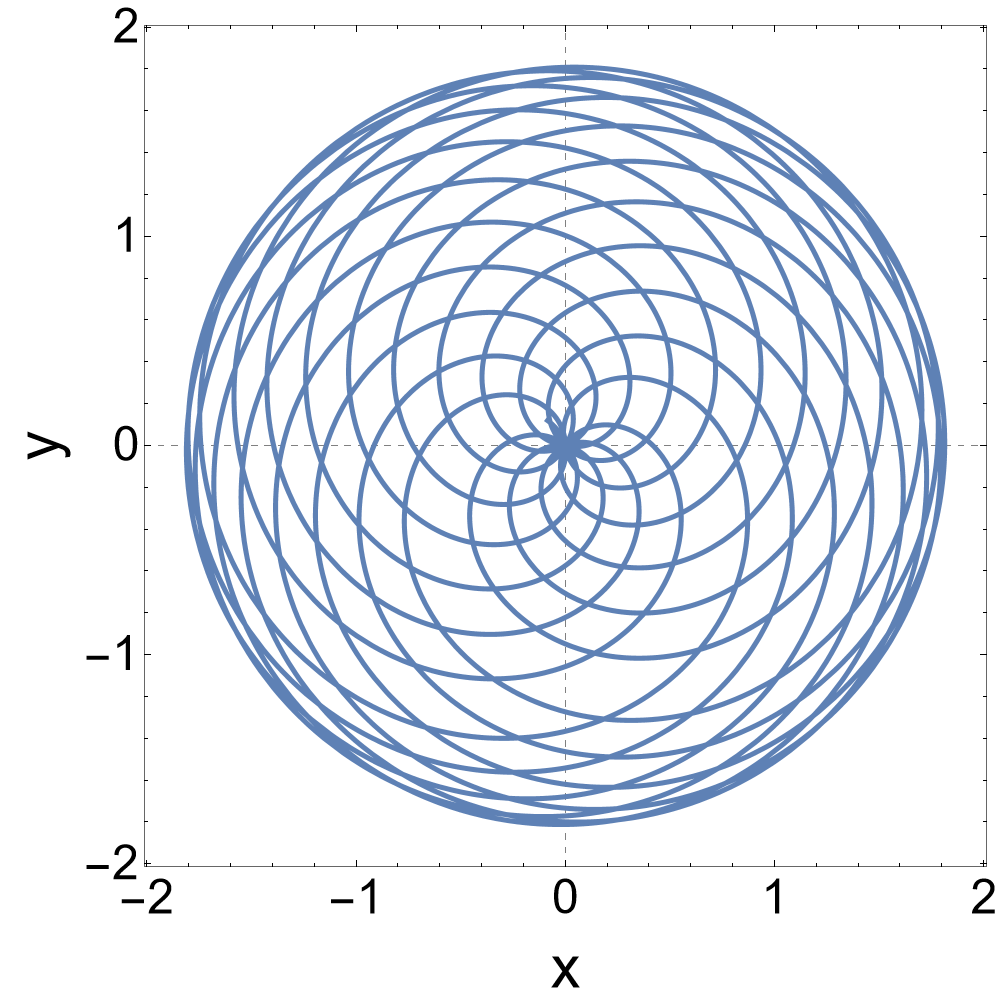}}
        \subfigure[$\chi=0$, $\Delta=-0.08$, $\tilde{\Delta}=0.56$.]{\includegraphics[scale=.34]{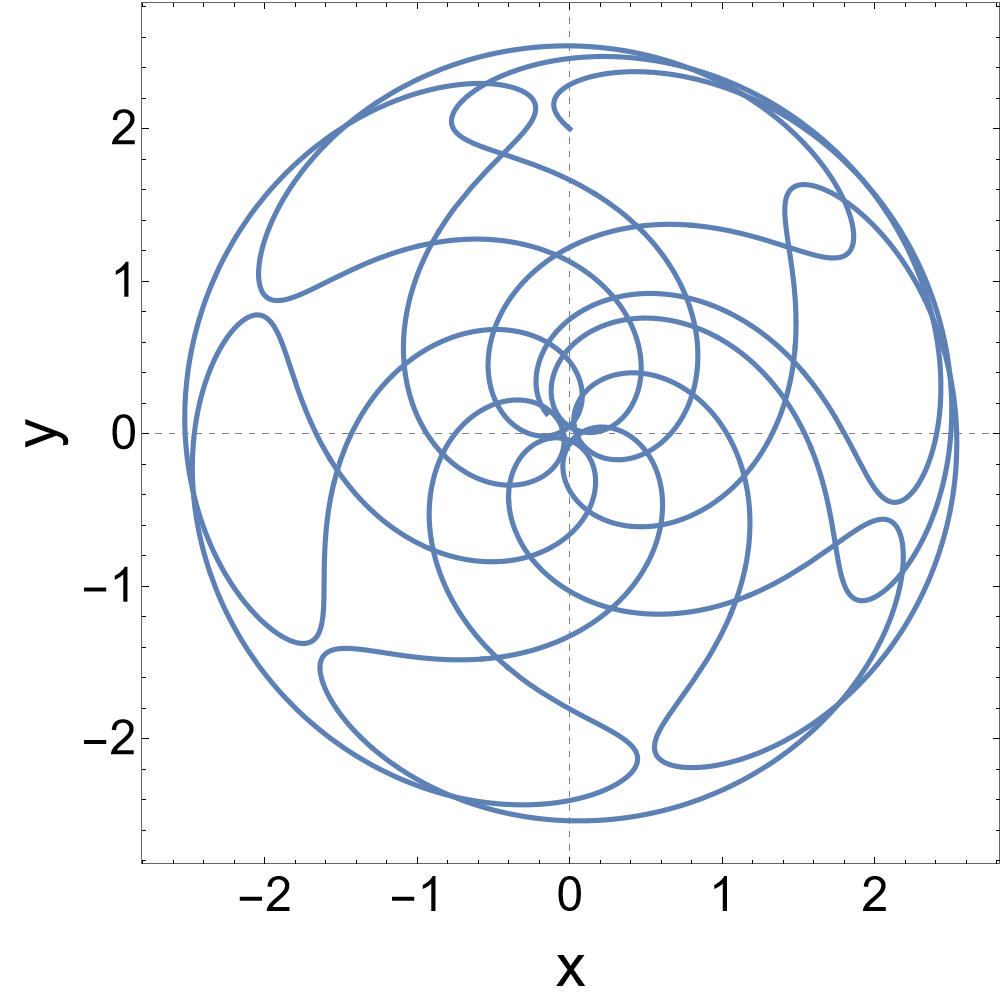}}
        \subfigure[$\chi=-1$, $\Delta = 0.26$, $\tilde{\Delta}= -0.32$.]{\includegraphics[scale=.34]{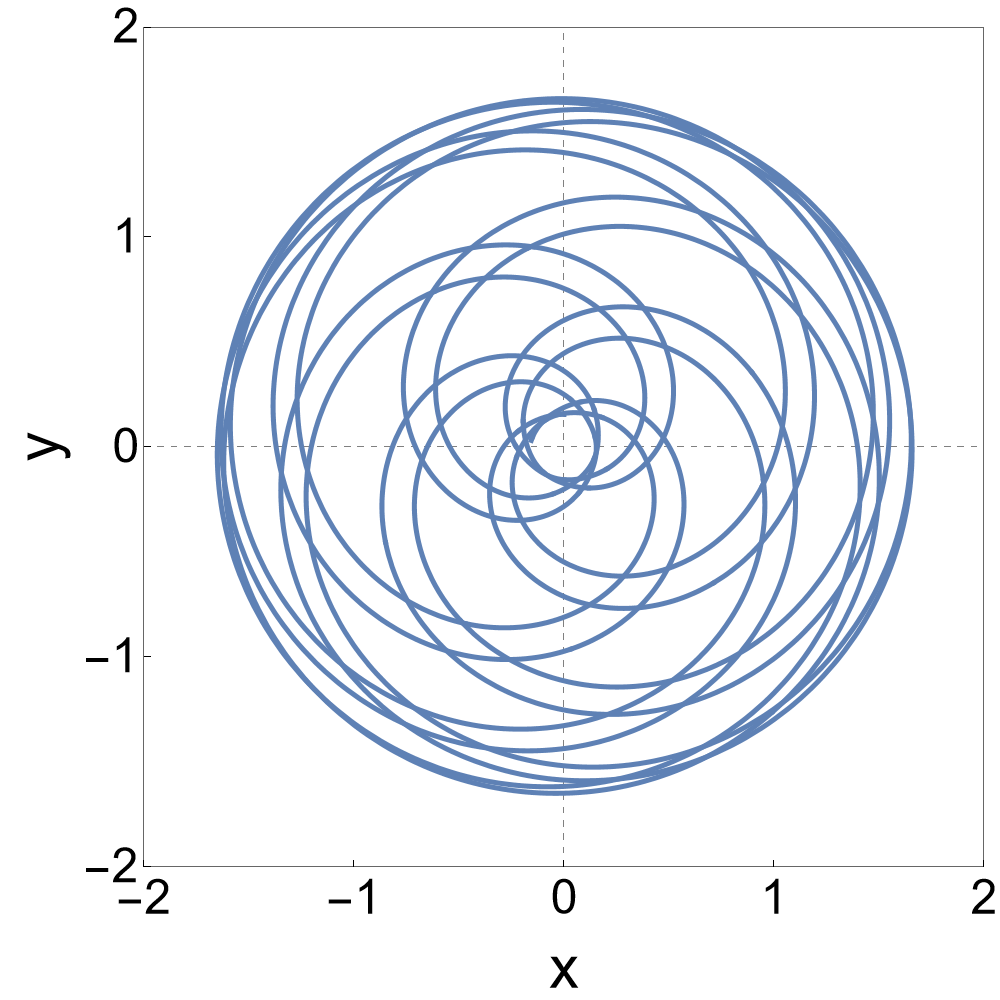}}
    \caption{Examples of planar orbits in the EMS space-time.  
    For orbits to be planar, i.e., for motion in a plane of constant $z$, the conditions $4jL+\Delta = 0$ and $\mathcal{E}^2 - k = 0$ must hold.
    All orbits shown are for $j=0.25$ and $N^2 = 0.25$. 
    In Figure (a) we show a radial escape orbit. 
    Such an orbit is only possible for uncharged massless particles with no angular momentum. 
    In Figure (b) we show an orbit similar to that in (a) for massive particles. 
    This orbit is bounded in the $\rho$-direction [see text]. 
    In Figure (c) we show the orbit of a charged massless particle with $L = \Delta = 0$, and in Figure (d) the one of a charged massive particle with $L \neq 0$.}
    \label{orbits1_planar_motion}
\end{figure}

First insight concerning the possible motion can be obtained directly from Eq.~(\ref{rho motion}). 
In order to exist, physical orbits must satisfy $R(\rho) \geq 0$.
This leads to the inequality
\begin{equation}
\label{classification - rho bound}
    k \geq \frac{\Bar{L}^2}{\rho^2} - \Lambda \chi.
\end{equation}
Note that $-\Lambda \chi \geq 0$ everywhere and hence there is a lower bound on $k$. 
On the other hand, from Eq.~\ref{zet motion} we have
\begin{equation}
    k \leq \left[ \mathcal{E} + \left( 4jL+\Delta \right)z \right]^2,
\end{equation}
which then puts an upper bound on $k$. Interestingly, for the geodesic motion of massless particles with no angular momentum, there are no turning points in the radial direction, and therefore the particles can escape to radial infinity.
In addition, if $\mathcal{E}^2 = k$ the orbit will lie in a plane with $z$ constant.
Such an orbit is shown in Fig.~\ref{orbits1_planar_motion}(a). 
On the other hand, for the geodesic motion of massive particles with no angular momentum, the combined bounds become 
\begin{equation}
 1 + 2N^2 \rho^2 + (j^2 + N^4)\rho^4 \leq k \leq \mathcal{E}^2 , 
\end{equation}
which thus puts no restrictions on the $z$-direction, but gives turning points in the $\rho-$ direction.
Therefore the possible orbits will be bounded in the $\rho$-direction. 
The motion in the $z$-direction can then be planar for $\mathcal{E}^2=k$, as shown in Fig.~\ref{orbits1_planar_motion}(b), or escape to infinity for $\mathcal{E}^2 > k$.

\begin{figure}[p]
    \centering
    \mbox{ \hspace*{-0.5cm}
    \subfigure[$\chi=0$, $\Delta = 0$, $\tilde{\Delta}=0$, $L=0$, $k=2$ and $\mathcal{E}=2.1$.]{\includegraphics[scale=0.3]{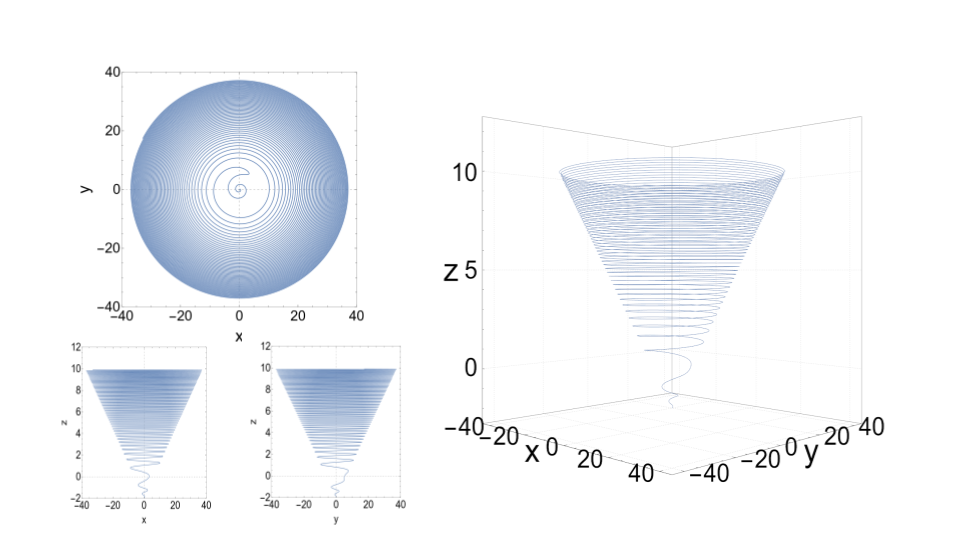}}
    \hspace*{-1.0cm}
    \subfigure[$\chi=-1$, $\Delta = 0$, $\tilde{\Delta}=0$, $L=0.6$, $k=6$ and $\mathcal{E}=\sqrt{6}$.]{\includegraphics[scale=0.3]{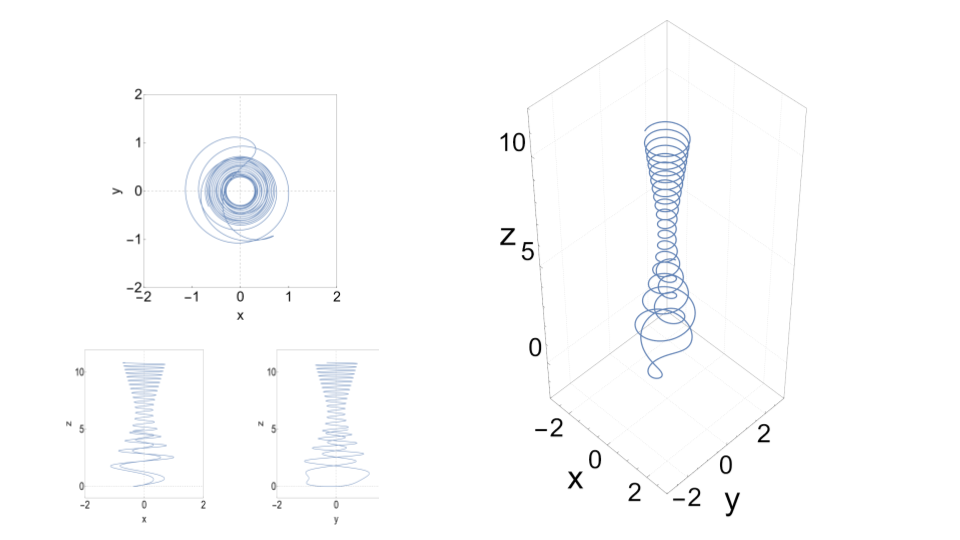}}
    }
    \mbox{ \hspace*{-0.5cm}
    \subfigure[$\chi=0$, $\Delta = 0.2$, $\tilde{\Delta}=-0.4$, $L=0.5$, $k=2$ and $\mathcal{E}=2$.]{\includegraphics[scale=0.3]{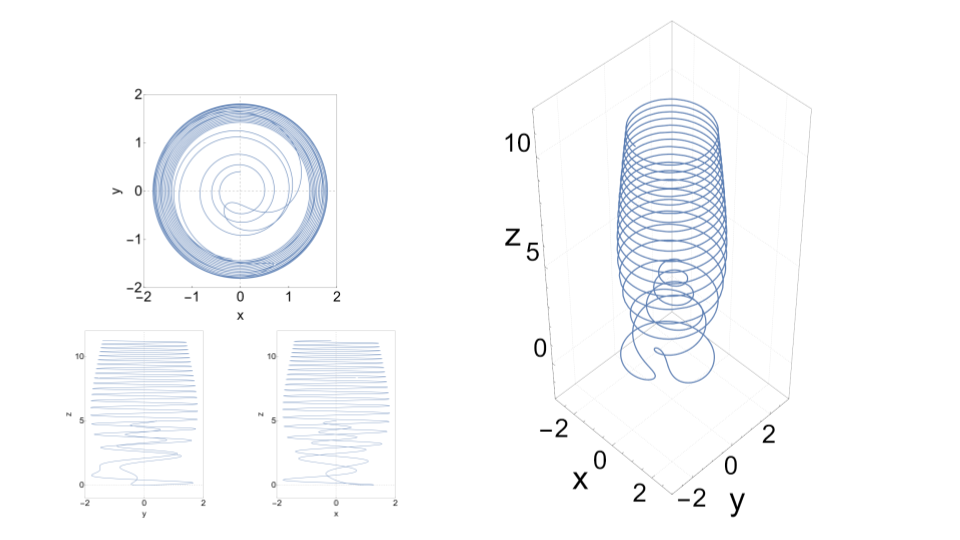}}
    \hspace*{-1.0cm}
    \subfigure[$\chi=-1$, $\Delta = -0.7$, $\tilde{\Delta}=0.3$, $L=0.5$, $k=4$ and $\mathcal{E}=2.1$.]{\includegraphics[scale=0.3]{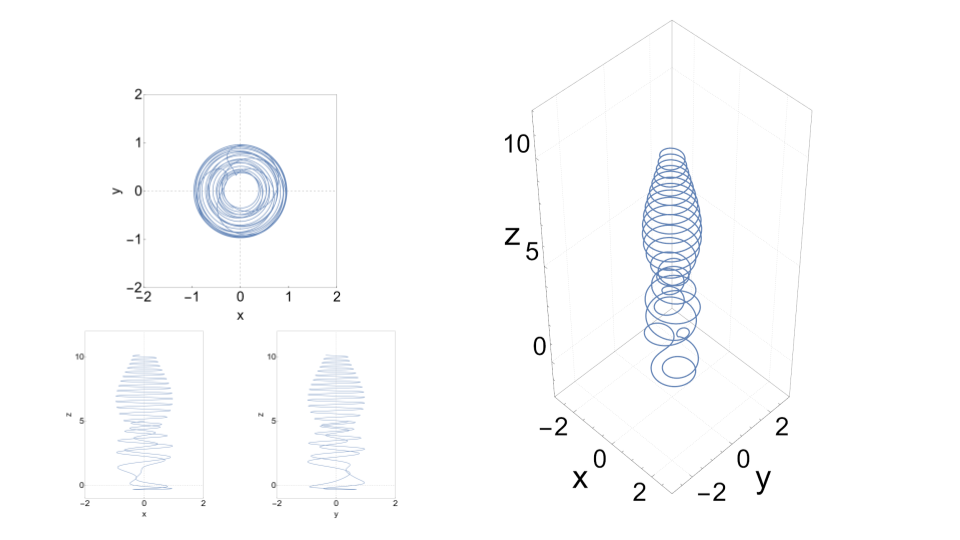}}
    }
    \caption{Examples of non-planar orbits in the EMS space-time. Non-planar orbits will always escape to infinity in the $z$-direction. 
    The test particles can either escape directly to infinity or reach a turning point in the $z$-direction and then escape to infinity. 
    For all orbits here, we have set $j=0.5$ and $N^2 = 0.5$. 
    In Figure (a) we show a geodesic orbit for a massless particle that escapes in $\rho$- as well as $z$-direction, with parameters chosen as those in Fig.~\ref{orbits1_planar_motion}(a) apart from the value of the energy. 
    In figure (b) we show a geodesic orbit for a massive particle with $L\neq 0$. 
    In figures (c) and (d) we show two orbits for charged particles, massless and massive, respectively.}
    \label{orbits2_non_planar}
\end{figure}

For uncharged particles with angular momentum, $L \neq 0$, or charged particles, the term $ \Bar{L}^2/\rho^2$ in Eq.~(\ref{classification - rho bound}) imposes an inner turning point in the $\rho-$direction, and therefore the allowed motion is even more restricted.
The physical orbits will then oscillate between these two turning points. 
Note that, for neutral particles with $L \neq 0$, the motion is necessarily non-planar.
However, for charged particles, it is possible to counter-balance this dragging effect.
Therefore, in the $z$-direction, the particle can either (i) move in a plane or (ii) escape to infinity.
In case (i)  $4jL + \Delta = 0$ and $\mathcal{E}^2 = k$.
Examples of such planar orbits are shown in Fig.~(\ref{orbits1_planar_motion})(c) 
for massless particles, and in (d) for massive particles. 
In case (ii) the particles can either escape directly to infinity or reach a turning point in the $z$-direction, and then escape to infinity, depending on the initial conditions.
Examples of such orbits are shown in Fig.~(\ref{orbits2_non_planar}) for massless and massive particles and for both geodesic motion and charged particle motion, respectively.

\section{Summary and Conclusions}
\label{conclusions}

The electromagnetic swirling universe is a novel solution constructed recently in \cite{Barrientos:2024pkt} by applying a composition of the Harrison and Ehlers transformations to a Minkowski seed solution.
It describes a space-time possessing electric and magnetic fields together with the swirling behavior. 
It can be considered an immersion of the Melvin universe into a swirling universe (or vice-versa). 
These two cases can be recovered as limiting cases, i.e., when either the swirling parameter vanishes or the electromagnetic field vanishes, respectively.

In this paper, we have focused on the motion of test particles in the EMS space-time.
Since this space-time is equipped with a U(1) gauge field, we considered the motion of charged particles, with both electric and magnetic charge, the latter for the sake of completeness. 
The equations of motion can be separated in the Hamilton-Jacobi formalism and can be analytically integrated in terms of elementary and elliptic functions. 
The solutions are completely characterized by the four constants of motion, namely the particle's total energy, its angular momentum with respect to the symmetry axis, the normalization condition, and the separation constant. 
The mathematical structure of the decoupled geodesic equations resembles the one in the swirling universe.

Insight into the type of test particle motion possible can be obtained directly from the equations of motion. 
A typical orbit, just like in the swirling universe, is bounded in the $\rho$-direction and escapes to infinity in the $z$-direction. 
Geodesic motion can be planar only in the absence of angular momentum.
In that case, a massless particle can escape to radial infinity. 
However, for charged particles, the electromagnetic interaction can counter-balance the dragging effect, and orbits can also be planar by a proper choice of angular momentum and energy.
This is possible for both types of particles.

The next natural step in our research would be to consider the immersion of a black hole into the EMS space-time. 
Similarly to the swirling, or the Melvin universe, the EMS space-time can be thought of as a background in which compact objects, such as black holes, can be immersed. 
The immersion of a Schwarzschild black hole has already been discussed in (\cite{Barrientos:2024pkt}).
However, the space-time containing a black hole is no longer of Petrov type D, but of the more general type I.
Therefore, a full separation of variables is not expected.

\section*{Acknowledgements}

R.C. would like to thank CAPES for financial support under Grant No: 88887.371717/2019-00. 
J.K. gratefully acknowledges support by DFG project Ku612/18-1.

\appendix

\section{Affine parameter}
\label{affine}

The motion discussed in the main body of the paper has been 
given in terms of the Mino time, $\lambda$, which relates to the affine parameter, $\tau$, via
\begin{equation}
    \frac{{\rm d} \tau}{{\rm d}\lambda} = \Lambda.
\end{equation}
This is an elliptic integral of the third kind and, hence, can be solved analytically. 
By performing the same transformations as discussed in Sec.~(\ref{section rho motion}), we find
\begin{equation}
    \frac{{\rm d} \tau}{{\rm d}\lambda} = \epsilon_0 + \frac{\epsilon_1}{v-\alpha} + \frac{\epsilon_2}{\left(v-\alpha \right)^2} \, 
\end{equation}
with $\alpha - b_2/12$, and the constants are
\begin{equation}
    \epsilon_0 = \Lambda\left( \sqrt{y_0} \right), 
    \ \ \ \qquad \ \ \ 
    \epsilon_1 = \frac{b_3 \left( N^2 + (j^2+N^4)y_0 \right)}{2},
    \ \ \ \qquad \ \ \
    \epsilon_2 = \frac{b_3^2  (j^2+N^4) }{16}.
\end{equation}
Thus, the solution is given by
\begin{equation}
    \tau(\lambda) = \epsilon_0 + \epsilon_1 \mathcal{I}_1(\lambda, y_\alpha) + \epsilon_2 \mathcal{I}_{2}(\lambda, y_\alpha).
\end{equation}

\section{Integration of elliptic integrals of the third kind}
\label{integration of elliptic integrals of the third kind}

Some of the integrals discussed in this paper are elliptic integrals of the third kind. 
In this appendix, we present the formulae to evaluate integrals of the type: 
$\mathcal{I}_{n} = \int_{v_0}^{v} \frac{1}{\left( \wp(v')-\gamma \right)^{n}}$, 
with $n=1$,$2$ or $3$, where $\gamma = \wp(y_\gamma)$ is a single pole of the above function. 
A table with these and other relations can be found in \cite{kerrweierstrass}.

The starting point is to expand the function as
\begin{equation}
\label{inverse wp expansion}
    \frac{\wp'(y)}{\wp(v)-\wp(y)} = \zeta\left( v - y \right) - \zeta\left( v + y \right) + 2 \zeta \left( y \right),
\end{equation}
which can then be directly integrated using the definition of the 
$\sigma(x)$-function, $\ln \sigma(x) = \int \zeta(x) {\rm d}x$, to get
\begin{equation}
\label{int I1}
        \mathcal{I}_{1}\left( v,y \right) = \int \frac{{\rm d}v}{\wp(v)-\wp(y)} = \frac{1}{\wp'(y)} \left[ 2 \zeta(y) v + \ln \frac{\sigma \left( v-y \right)}{\sigma \left( v+y \right)} \right].
\end{equation}  

Terms of higher order can be found after performing some algebra on Eq.~(\ref{inverse wp expansion}). Taking the derivative of (\ref{inverse wp expansion}) with respect to $y$, and using $\zeta(x) = - \int \wp(x) {\rm d}x$ one gets 

\begin{equation}
    \frac{1}{\left(\wp(v)-\wp(y)\right)^2} = \frac{1}{\wp'(y)^2} \left[ \wp(v-y) + \wp(v+y) +2\wp(y) - \frac{\wp''(y)}{\wp(v)-\wp(y)} \right] .
\end{equation}
This can then be directly integrated and lead to 
\begin{equation}
\label{int I2}
        \mathcal{I}_{2} \left( v,y \right)= \int \frac{{\rm d}v}{\left( \wp(v)-\wp(y) \right)^2} =-\frac{\wp''(y)}{\wp'(y)^2} \mathcal{I}_1 - \frac{1}{\wp'(y)^2} \left( \zeta \left( v+y \right) + \zeta \left( v-y \right) + 2\wp(y) v \right).
\end{equation}
Applying the same idea and taking the second derivative of Eq.~(\ref{inverse wp expansion}) with respect to $y$, one gets
\begin{eqnarray}
     \frac{1}{\left( \wp(v)-\wp(y) \right)^3} = \frac{1}{2\wp(y)^3} \left[ \wp'(v-y) + \wp'(v+y) - 2\wp'(y) -\frac{12 \wp'(y)\wp(y)}{\wp(x)-\wp(y)} - \frac{3 \wp'(y)\wp''(y)}{\left( \wp(v)-\wp(y) \right)^2} \right],
\end{eqnarray}
where $\wp^{(3)}(y) = 12 \wp(y) \wp'(y)$ has been used. 
Integrating the above expression then leads to
\begin{eqnarray}
     \mathcal{I}_{3} \left( v,y \right)& = &\int \frac{{\rm d}v}{\left( \wp(v)-\wp(y) \right)^3} \nonumber \\
     &=&\frac{1}{2\wp(y)^3} \left[ \wp(v+y) - \wp(v-y) - 2\wp'(y) v -12 \wp'(y)\wp(y) \mathcal{I}_1 - 3 \wp'(y)\wp''(y)\mathcal{I}_2 \right].
\end{eqnarray}

\end{document}